\documentclass[12pt,letterpaper]{article}

\usepackage{textcomp}
\usepackage{fullpage}
\usepackage{amsfonts}
\usepackage{verbatim}
\usepackage[francais,english]{babel}
\usepackage[utf8]{inputenc}
\usepackage[T1]{fontenc}
\usepackage{pifont}
\usepackage{color}
\usepackage{setspace}
\usepackage{lscape}
\usepackage{indentfirst}
\usepackage{tabularx}
\usepackage[normalem]{ulem}
\usepackage{booktabs}
\usepackage{natbib}
\usepackage{float}
\usepackage{latexsym}
\usepackage{url}
\usepackage[hidelinks]{hyperref}
\usepackage{epsfig}
\usepackage{graphicx}
\usepackage{amssymb}
\usepackage{amsmath}
\usepackage{bm}
\usepackage{array}
\usepackage{mhchem}
\usepackage{ifthen}
\usepackage{caption}
\usepackage{hyperref}
\usepackage{transparent}
\usepackage{amsthm}
\usepackage{amstext}
\usepackage{tabu}
\usepackage[utf8]{inputenc} 
\usepackage{natbib}
\usepackage{pdfpages}

\renewcommand{\section}[1]{%
\bigskip
\begin{center}
\begin{Large}
\normalfont\scshape #1
\medskip
\end{Large}
\end{center}}

\renewcommand{\subsection}[1]{%
\bigskip
\begin{center}
\begin{large}
\normalfont\itshape #1
\end{large}
\end{center}}

\renewcommand{\subsubsection}[1]{%
\vspace{2ex}
\noindent
\textit{#1.}---}

\renewcommand{\tableofcontents}{}

\bibpunct{(}{)}{;}{a}{}{,}  

\begin{document}
\begin{flushright}
Version dated: \today
\end{flushright}
\bigskip

\bigskip

\medskip
\begin{center} 
\noindent{Detecting patterns of species diversification in the presence of both rate shifts and mass extinctions}
\bigskip



\noindent {\normalsize \sc Sacha Laurent$^{12}$, Marc Robinson-Rechavi$^{12}$, and Nicolas Salamin$^{12}$}\\
\noindent {\small \it 
$^1$Department of Ecology and Evolution, University of Lausanne, 1015 Lausanne, Switzerland\\
$^2$Swiss Institute of Bioinformatics, Quartier Sorge, 1015 Lausanne, Switzerland}\\
\end{center}
\medskip





\noindent (Keywords: Phylogeny, Diversification, Mass Extinctions)\\

\section{Abstract}
\textbf{Background:} Recent methodological advances allow better examination of speciation and extinction processes and patterns. A major open question is the origin of large discrepancies in species number between groups of the same age. Existing frameworks to model this diversity either focus on changes between lineages, neglecting global effects such as mass extinctions, or focus on changes over time which would affect all lineages. Yet it seems probable that both lineages differences and mass extinctions affect the same groups.\\
\textbf{Results:} Here we used simulations to test the performance of two widely used methods under complex scenarios of diversification. We report good performances, although with a tendency to over-predict events with increasing complexity of the scenario. \\
\textbf{Conclusion:} Overall, we find that lineage shifts are better detected than mass extinctions. This work has significance to assess the methods currently used to estimate changes in diversification using phylogenetic trees. Our results also point toward the need to develop new models of diversification to expand our capabilities to analyse realistic and complex evolutionary scenarios.


\section{Background}
The estimation of the rates of speciation and extinction provides important information on the macro-evolutionary processes shaping biodiversity through time \citep{Ricklefs2007}. Since the seminal paper by Nee $et$ $al.$ \cite{Nee1994}, much work has been done to extend the applicability of the birth-death process, which now allows us to test a wide range of hypotheses on the dynamics of the diversification process.\\

Several approaches have been developed to identify the changes in rates of diversification occurring along a phylogenetic tree. Among them, we can distinguish between lineage-dependent, trait-dependent, time-dependent and diversity-dependent changes. Lineage specific methods identify changes in macro-evolutionary rates ~--- speciation and extinction rates, denoted as $\lambda$ and $\mu$, respectively ---~ at inner nodes of a phylogenetic tree \citep{Rabosky2007a, Alfaro2009, Silvestro2011}. We can also identify trait-dependence in speciation and extinction rates if the states of the particular trait of interest are known for the species under study \citep{Maddison2007, FitzJohn2009, Mayrose2011}. It is also possible to look for concerted changes in rates on independent branches of the phylogenetic tree by dividing it into time slices \citep{Stadler2011}. Finally, diversity-dependent effects can be detected when changes of diversification are correlated with overall species number \citep{Etienne2012}. Most methods can correct for incomplete taxon sampling, by assigning species numbers at tips of the phylogeny \citep{Alfaro2009, Stadler2013}, or by introducing a sampling parameter \citep{Nee1994}. By taking into account this sampling parameter at time points in the past, it is also possible to look for events of mass extinction \citep{Stadler2011}.\\

These methods provide insights into the dynamics of species diversification and it is now well accepted that differences in lineage-specific rates exist \citep{Jetz2012, Barker2013}. However, it seems unlikely that both lineage specific shifts and mass extinction events would not have occurred, especially when studying large phylogenetic trees covering hundreds of million years of evolution. For example, several global crises, which caused the extinction of a high proportion of species \citep{Raup1982}, have occurred since the appearance of the last common ancestor of vertebrates. Among them, the Cretaceaous-Paleogene (K-Pg) boundary and the Permian-Triassic events, which happened $65$ million years ago (Mya) and $251$ Mya, respectively, induced the most dramatic losses of biodiversity \citep{Erwin2006}. Moreover, other less extensive events have also occurred in the past hundred million years \citep{Benton1995}.\\

Alternative models have been proposed for mass extinctions. They could be represented as a high number of species disappearing at the same time (single-pulse model), or as an increase of the background rate of extinction during an extended period of time (time-slice model) \citep{Condamine2013}. They could also impact biodiversity in different ways. Three main hypotheses, corresponding to different patterns of extinction, have been proposed \citep{Raup1992}. First, the event could affect all lineages equally and terminate any extant lineage with the same probability. This "field of bullets" scenario is often used as a null model \citep{Nee1997, Faller2008}. Second, in the "fair game" scenario, some form of lineage selection would occur, where the most successful species ~--- in our case, the most diversifying species ---~ before the event would be the most likely to survive. This could, for instance, happen if the probability of survival depends on a specific trait varying across the lineages of the phylogeny \citep{Faller2012}. Finally, in the "wanton destruction" scenario \citep{Eble1999}, the event could induce such changes in the environmental conditions that the probability of extinction of the species and their post-event diversification rate would be uncorrelated to their initial speciation and extinction rates.\\

Although lineage-dependent differences in macro-evolutionary rates and mass extinctions are known to happen, the performances of the existing methods to identify both lineage-specific rate shifts when mass extinctions have occurred, and mass extinctions when lineage-specific rate shifts have occurred has not, to our knowledge, been investigated. The aim of this study was thus to assess the performance of current methods to estimate the rates of diversification using complex scenarios involving both mass extinctions and lineage shifts. We used simulations to assess the impact of varying number and magnitude of rate shifts and mass extinction events.\\

\section{Methods}

\begin{figure}[h!]
\begin{center}
\includegraphics{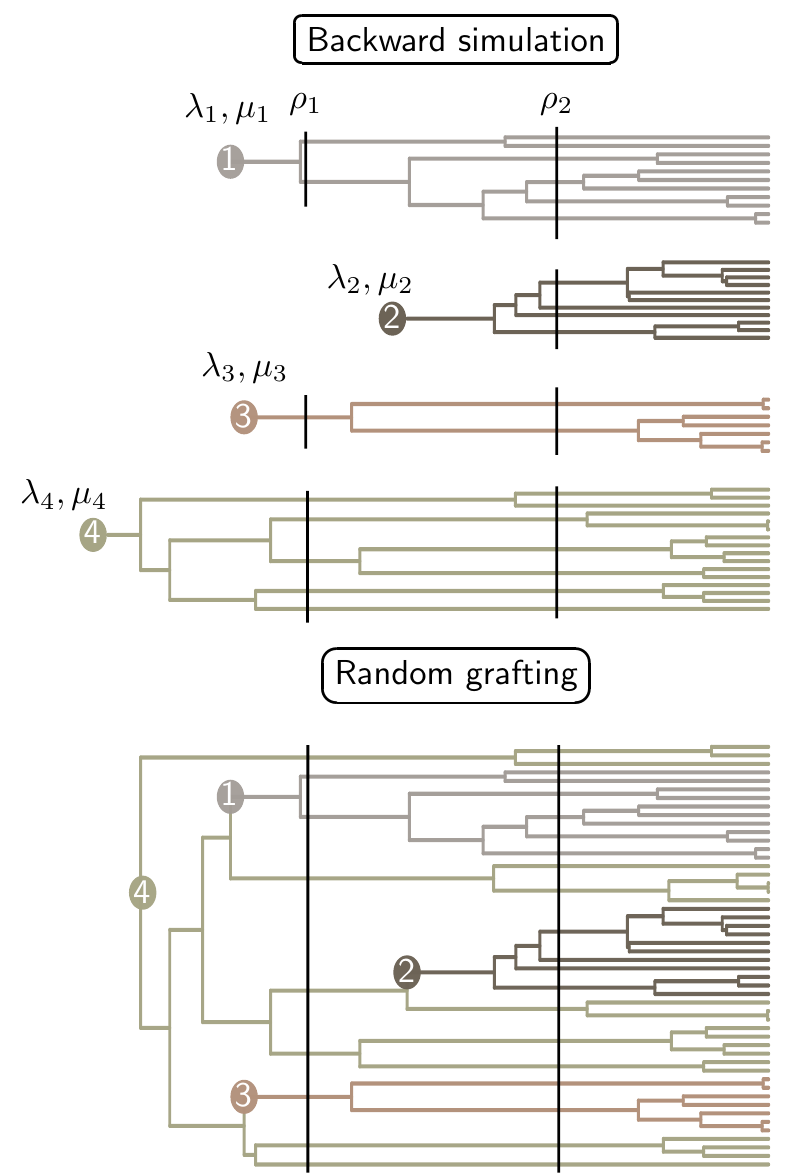}
\end{center}
\caption{Workflow of the simulation process. Hypothetic case of $50$ species tree, $3$ lineages shifts and $2$ mass extinctions. The number of species in each lineage is randomly drawn first. Each tree is grown separately with different $(\lambda, \mu)$ but with identical survival rates ($\rho$) at each mass extinction events. The four trees are then successively joined at branches ensuring ultrametricity. Vertical continuous lines: simulated mass extinction events, full circles: ancestor where diversification change occurred.}
\label{workflow}
\end{figure}

Figure \ref{workflow} gives an overview of the simulation design. We used a backward algorithm to simulate phylogenetic trees as implemented in the function $sim.rateshift.taxa$ from the R \citep{R} package TreeSim \citep{Stadler2011d}. Direct forward approaches to simulate trees using a birth-death process are also available. They can be used by conditioning either on the number of tips or on the total amount of time of the process. The former approach can lead to bias \citep{Hartmann2010}, while the latter could be less practical in our specific context as the procedure would result in trees with highly variable numbers of taxa, in particular when adding mass extinction events. A backward simulation procedure is therefore the best solution to simulate the different diversification scenarios of interest for our study. This procedure enables both single-pulse or time-slice modeling of mass extinctions, but we chose to represent them only using the single-pulse model because paleontological data indicates very high species loss at major mass extinction events in a limited amount of time. For instance, a 52\% decrease in marine families was observed at the Permian-Triassic boundary \citep{Raup1982}.\\

\begin{figure}[h!]
\begin{center}
\includegraphics{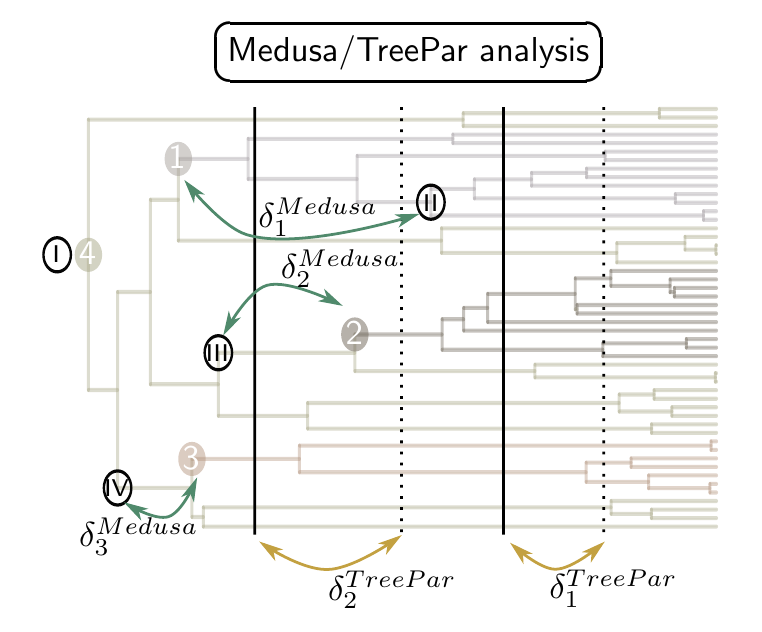}
\end{center}
\caption{Exemple output of the analyzes. We run the Medusa and TreePar analysis, and group the pairs of simulated/estimated events by minimizing the sum of the distance separating the events in each pair ( $\sum \limits_i \delta_i^{Medusa}$ and $\sum \limits_i \delta_i^{TreePar}$ ). Vertical dotted lines: estimated mass extinction events by TreePar, dotted circles with roman letters: estimated diversification rate shift by Medusa, by decreasing significance, other: as in figure \ref{workflow}. The first estimated shift is always at the root of the tree.}
\label{analysis}
\end{figure}

Our algorithm takes as input the number of extant species, the evolutionary rates $\lambda$ and $\mu$, and the time of occurrence and survival rate $\rho$ for mass extinction events. We assumed in the first part of our simulations that these events happened according to the field of bullet scenario (step $1$). We randomly grafted different trees having experienced the same mass extinction events but different evolutionary rates to account for rate shifts in diversification (step $2$; see Table \ref{table1}). First, we ran as many backward simulations as the number of lineages shifts in our tree. We defined the number of species in each backward simulation by drawing samples from a Dirichlet distribution to keep the total sum equal to the overall number of species. We then ranked the trees by decreasing order of their total age, which included the stem branch length provided by TreeSim. We selected from the oldest tree (referred to as acceptor tree) the branches that overlapped in time with the age of the stem branch of the second oldest tree (referred to as donor tree). Thus, the branches considered for possible grafting were the ones that included the age of the donor tree between the timing of the two speciation events defining them in the acceptor tree. We randomly chose one of those branches to graft the donor tree onto the acceptor. This ensures ultrametricity of the newly created tree and leaves the branch lengths of each separate tree unmodified once the lineage having experienced the diversification shift is removed. We iterated over this protocol until all donor trees, whose number varied in our simulations between 0 and 5 (Table \ref{table1}), were grafted. Finally, we ran Medusa \citep{Alfaro2009} and TreePar \citep{Stadler2011} analyses on each simulated tree to investigate our capacity to recover the signal of mass extinctions and diversification shifts (Fig. \ref{analysis}).  We simulated trees with different numbers of lineages and extinction events to assess the influence of these factors. Table \ref{table1} summarizes the parameter space explored for the $16,371$ trees that we simulated. For the values of $\lambda$ and $\mu$, we targeted distributions similar to the estimates calculated on a mammalian phylogeny \citep{Bininda-Emonds2007}.\\

\tabulinesep=1.1mm
\begin{table}
\begin{center}
\begin{tabu}{lc}

\bf{Parameter} & \bf{Possible values}\\
\hline
$\lambda$ & $Unif(0.05, 0.25)$ \\
$\mu$ & $Unif(0, 0.05)$\\
$\rho$& $Unif(0.2, 0.9)$\\
Number of tips & $200, 500, 1000, 2500, 5000$\\
Mass extinction event number & $0$ to $5$ \\
Rateshift event number & $0$ to $5$\\
Mass extinction event time& $ Unif(0, min(\frac{Log(N_i)}{\lambda_i-\mu_i}))$\\
\hline
\end{tabu}
\caption{Universe explored for parameters values. $Unif$: Uniform distribution, $i$: lineage identifier}
\label{table1}
\end{center}
\end{table}

Despite the issues to use existing forward algorithm, we nevertheless compared our backward algorithm with a ``forward-like'' algorithm based on the R package TESS \citep{Hohna2013}. We simulated trees with different values of $\lambda$, $\mu$ and species number to model the lineage shifts in diversification rates. We carry a similar grafting process as described in our backward algorithm. However, we removed all daughter species of the sister clade of the donnor tree in the acceptor tree. This step has the consequence of removing the instance of the artificially created speciation event that was present in our former algorithm and effectively mimic a forward algorithm with a change in diversification rate possible anywhere between two speciation events. As the first conditioning is made on the number of species, and as we subsequently remove species, the total number of species at the end of the process in not constant but varies slightly below the number used for the conditioning. We simulated trees according to both our backward and forward algorithms and compared them using two different measurements: the distribution of branching times and the outcomes of Medusa on both our trees (Additional file 1). These two measures resulted in very similar outcomes and we present here only the results obtained by the backward algorithm.\\

Medusa is a maximum likelihood-based framework to detect shifts in diversification by iteratively adding breakpoints on inner branches of the tree with different rates of speciation and extinction. It uses $\Delta$AIC to discriminate between models with an increasing number of parameters \citep{Alfaro2009}. Rabosky also recently presented a new method (BAMM) to estimate the number of possible rate changes along a phylogenetic tree and to fit exponential responses in macroevolutionary rates to time or to species number \cite{Rabosky2014a}. Unlike Medusa, BAMM uses a Bayesian framework, with reversible jump Markov chain Monte Carlo to estimate the number of shifts in diversification in the phylogeny. In our design, we chose not to simulate varying speciation and extinction rates except at speciation nodes, thus using higher complexity models is not necessary. Comparisons between BAMM and Medusa have been performed, but only on simulations involving either time-dependent or diversity-dependent rates \citep{Rabosky2014a}. This framework led to a clear bias in favor of BAMM as Medusa can not evaluate such models, and resulted in Medusa estimating a lower number of events than what was actually simulated \citep{Rabosky2014a}. The numbers of estimated shifts obtained with Medusa can therefore be considered as conservative. Finally, we do not expect a different behavior for Medusa and BAMM regarding the identification of mass extinction events, as neither method incorporates them in their model. Those reasons, as well as the large computational burden to run Bayesian analyses on over $16,000$ trees, led us to favor the simpler Medusa framework for the rest of the study. Medusa was run until a more complex model was not supported by the $\Delta$AIC. We did not extract the macro evolutionary rate estimations from Medusa as we were only interested in testing the ability of the method to detect the events, and not the accuracy of the parameter estimation.\\

TreePar uses the birth-death process to identify changes in $\lambda$ and $\mu$ through time. This is done by estimating the probability of a change in parameter values within small time intervals, which can be extended to test for the occurrence of mass extinction events \citep{Stadler2011}. The parameters of the rate shifts might be correlated with those related to mass extinction \citep{Stadler2011}, which will be a problem for our simulations. We therefore restricted our analysis to the identification of mass extinction events to avoid this issue. The number of iterations of TreePar was set to the simulated number of mass extinction events plus one to test for the appearance of false positive events. A standard Likelihood Ratio Test (LRT) is used to extract the most likely models from TreePar and more complex models were favored when their p-value was less than $0.01$, following the standard approach for this framework \citep{Stadler2011}. Similarly to what was done with Medusa, we did not analyze estimations of survival rates at mass extinctions events given by this framework.\\

To verify that our simulation design had no effects on the methods evaluated, we tested the influence of the subtree grafting approach with a constant rate of diversification. We simulated trees with 200 species using both the standard procedures implemented in TreeSim and by grafting two subtrees of 150 and 50 species having evolved under the same $\lambda$ and $\mu$ values. We then compared the results obtained by TreePar and Medusa. We ran 250 pairs of simulations and we observed no significant differences in the number of false positive found between the groups with and without artificial grafting (7 and 13 for Medusa respectively, and none in both cases for TreePar), showing that our simulation design does not bias the estimation of the rate shifts by the two methods used.\\ 

We used a slightly different framework to study the impact of the different types of mass extinction events. We simulated a scenario that aimed at testing for the presence of the K-Pg mass extinction event using high order phylogenetic trees. We therefore simulated trees with a large number of extant species ($5,000$ tips, similar to the number of mammalian species) and a large number of lineage shifts ($5$), but only one event of mass extinction. The other parameters were still drawn at random from the ranges specified in Table \ref{table1}, except for the survival rate $\rho$ that was modified according to the models of mass extinction. For the fair game hypothesis, we randomly drew $\lambda$ and $\mu$ for the $5$ different lineage shifts, but the survival rate $\rho$ was modified for each lineage based on its diversification rate ($r$, $\lambda-\mu$). We thus considered that the trait influencing the probability of extinction for each species was its diversification rate. For the wanton destruction hypothesis, the mass extinction event induced a change in rates for each lineage, again drawn according to the distribution stated in Table \ref{table1}, and their survival rate $\rho$ was then based on their new diversification value. For the wanton destruction, our simulations included both a global rate shift and a mass extinction and we ran TreePar twice in order to detect both events. For the two latter cases, we chose to linearly parametrize $\rho$ with regards to diversification. As diversification could range between $0$ and $0.25$ and $\rho$ between $0$ and $1$, we applied a factor four to the diversification to obtain the survival rates of the lineages. We also ran Medusa on the three sets of simulations to assess the potential impact of the three extinction hypotheses on the detection of lineage shifts. For this second part, we generated over $700$ trees for each model of mass extinction event, for a total of $2289$ simulations.\\

\section{Results and discussion}
\subsection{Baseline performances}

\begin{figure}[h!]
\includegraphics{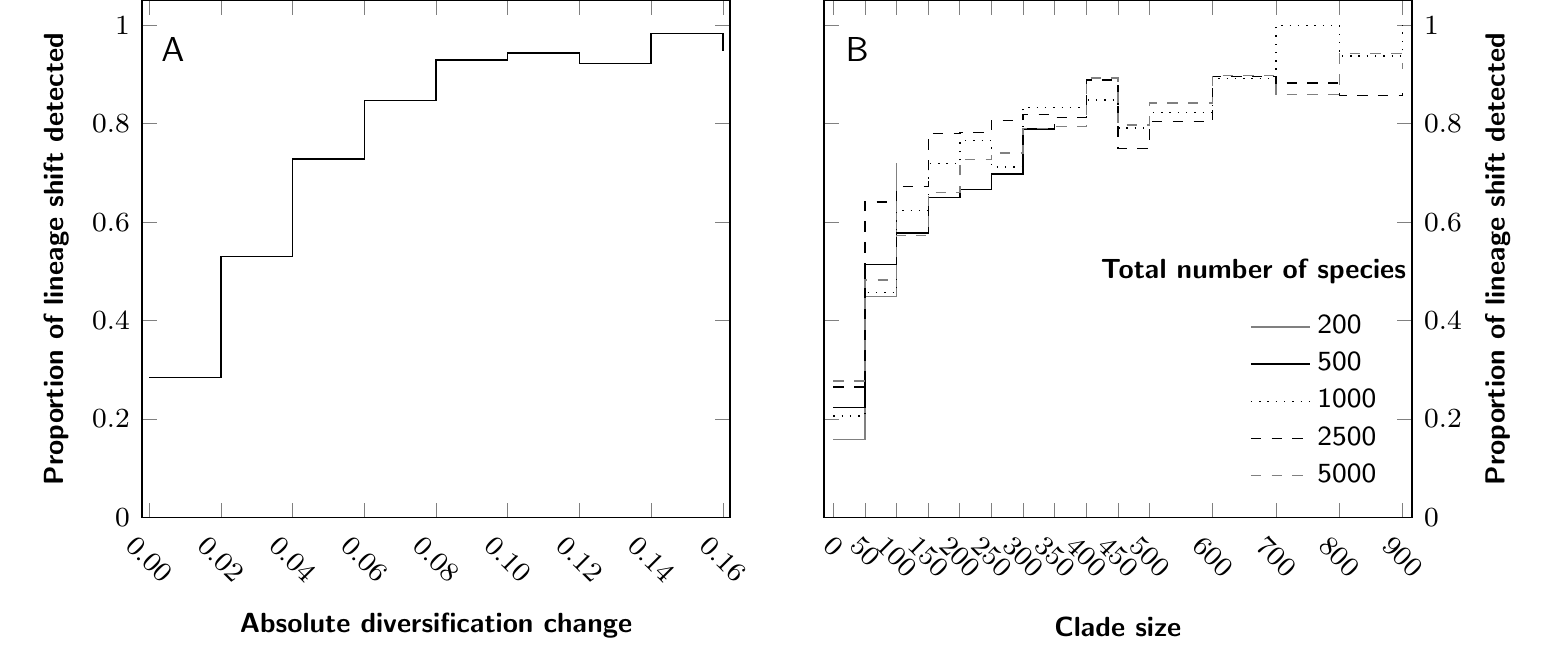}
\caption{Baseline detection level for Medusa, for simulations with one rate shift and no mass extinction event. A: Proportion of detected events for ranges of values of diversification, B: Proportion of detected events for ranges of extant species number in lineages.}
\label{Medusa_baseline}
\end{figure}

The backward and ``forward-like'' algorithms gave very similar results (Additional file 1) and we only present here the results obtained with the backward algorithm. To estimate the baseline behavior of both frameworks, we first tested the performance of the methods on the simplest scenarios. We thus selected simulations that included a single rate shift for Medusa, or a single mass extinction for TreePar. Figure \ref{Medusa_baseline} represents the fraction of shifts detected by Medusa relative to the absolute difference between the new and the old diversification values (Figure \ref{Medusa_baseline}A) and to the number of species in the lineage (Figure \ref{Medusa_baseline}B). More than $80\%$ of the changes in diversification larger than $0.05$ are detected by Medusa, which shows a good performance in assessing strong shifts. Further, Figure \ref{Medusa_baseline}B shows that the overall tree size has no influence on the detection, since lineages of the same size are as likely to be detected in small or larger trees. \\

\begin{figure}[h!]
\includegraphics{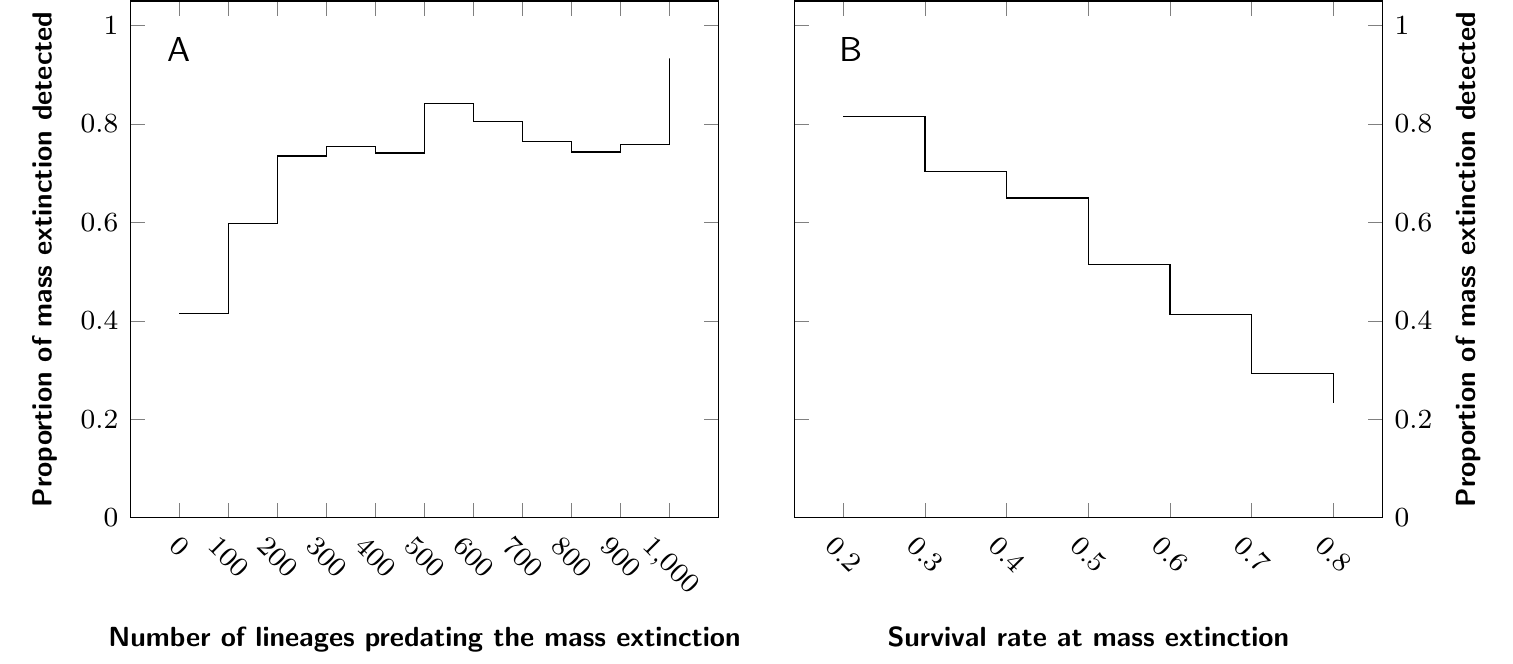}
\caption{Baseline detection level for TreePar, for simulations with one mass extinction and no diversification shift. A: Number of lineages predating the mass extinction event influence, B: Survival rate influence.}
\label{tp_baseline}
\end{figure}

We then checked the ability of TreePar to detect mass extinction as a function of the survival rate, $\rho$, as well as of the number of ancestral species predating this event in the reconstructed tree. We also used first the simplest simulation to limit the effect of other parameters. Figure \ref{tp_baseline}A shows that the signal of mass extinction in the phylogenetic tree is very weak when less than $100$ ancestral species are present before the event. This has implications for our ability to find evidence for the K-Pg boundary using phylogenetic trees of vertebrates, for example. We can only reach more than a hundred ancestral species older than $65$ My by considering phylogenetic trees encompassing distantly related lineages of tetrapods (see \cite{Bininda-Emonds2007} or \cite{Meredith2011}). Besides, as detection drops with increasing survival rate (Fig. \ref{tp_baseline}B), the signal is even less likely to be picked as the ancestors of the extant species might have experienced the mildest extinction rates.\\

\subsection{Mixed scenarios of diversification}

In a second stage, we analyzed simulations with more events and a mix of different types of events. We evaluated the performance of rate shift detection by Medusa, or of mass extinction events by TreePar, by comparing the events detected to the relevant simulated events. To perform the assignment between detected and simulated events (see Fig. \ref{analysis}), we chose to minimize the sum of the distances between each potential pairing of events ($\sum \limits_i \delta_i^{Medusa}$  and $\sum \limits_i \delta_i^{TreePar}$). The distance metric used for Medusa was the sum of the branch lengths along the shortest path separating the two nodes, whereas we used the time between the estimated and simulated pairs of mass extinction events for TreePar (see caption of Figure \ref{analysis} for details).\\

The simulations incorporated several factors and we tested the effect on the framework of three categorical parameters: total number of tips, number of mass extinctions and number of shifts in diversification rate (see Table \ref{table1} for their possible values). To ensure that the effects observed were related to the parameter of interest, we designed a reshuffling scheme for each parameter. First, we randomly selected an equal number of simulations for each combination of every possible value of the other two parameters. As an example, to study the outputs for trees of $200$ tips, we randomly drew an equal number of simulations with (i) no lineage shift, no mass extinction and $200$ tips; (ii) one lineage shift, no mass extinction and $200$ tips; (iii) one lineage shift, one mass extinction and $200$ tips; etc. This draw was repeated a hundred times and we determined, for each bin created, the proportion of simulations for which each method favored the model with the correct number of relevant events it was looking for, and the proportion of simulations for which they favored a model with too many events. Finally, we report the median and $95\%$ intervals of those proportions based on our hundred bins.\\

\subsubsection{Tree size influence}
\begin{figure}[h!]
\includegraphics{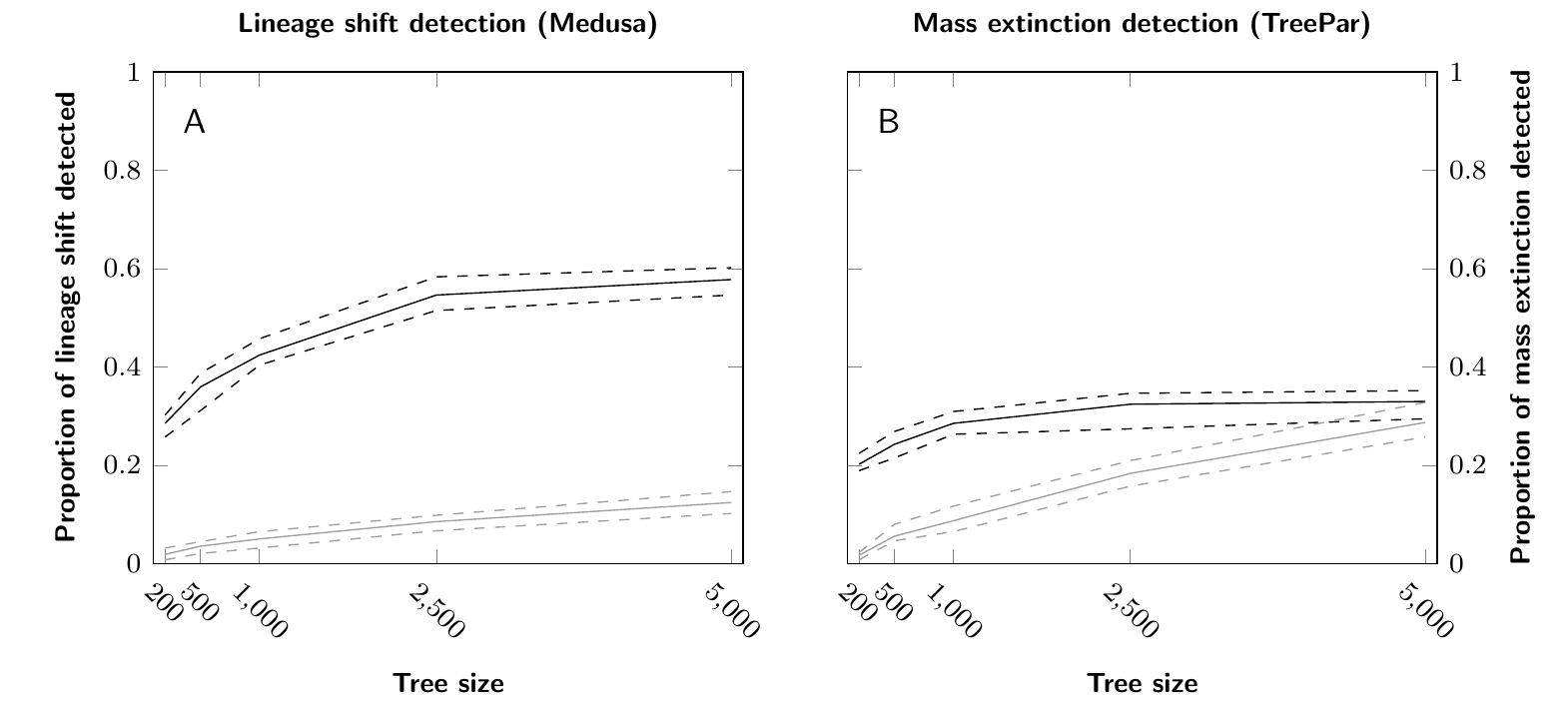}
\caption{Influence of tree size on the detection of lineage shifts (A) and mass extinction events (B). Continuous lines correspond to median proportions of simulations and dotted lines correspond to 95\% confidence interval, both based on resampling. Dark lines represent the proportion of simulations where the model with the correct number of events was the most favored, and light lines where a model with too many events was favored.}
\label{tree_size}
\end{figure}

\begin{figure}[h!]
 \begin{center}
\includegraphics{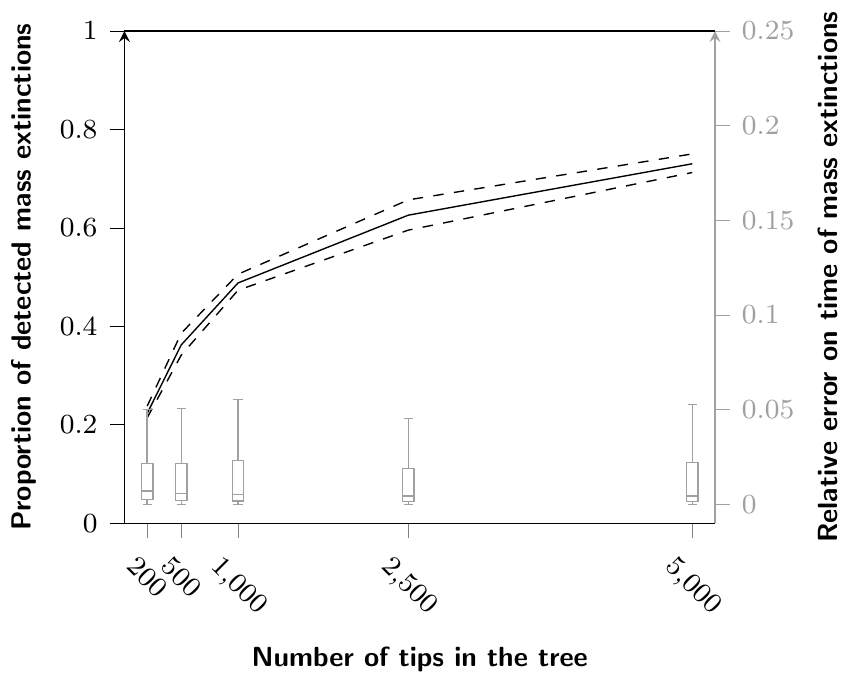}
\end{center}
\caption{Influence of tree size on the detection of mass extinctions by TreePar. Line: proportion of detected mass extinctions; boxplots: distribution of the errors on their timing relative to the time of the first speciation event of the tree.}
\label{tree_size_distance}
\end{figure}

Both Medusa and TreePar perform better in assessing the correct number of events they are set to detect with an increasing number of tips (Fig. \ref{tree_size}). The median proportion of simulations correctly assessed reaches $60\%$ for Medusa and $32\%$ for TreePar with $5,000$ tips. The increase in the number of tips also leads to an increased acceptance by TreePar of models with too many mass extinctions ($28\%$ for $5,000$ tips). However, the number of tips in the tree has no effect on the error of the estimated time of mass extinction (Fig. \ref{tree_size_distance}), even though more events are predicted. We only see a slight effect of tree size for Medusa, which is probably due to the fact that the method only detects lineage related events and does not depend on the total number of tips. We also investigated the effect of lineage size on the outputs of Medusa. We first compared the variance of lineage sizes relative to the overall tree size, contrasting the simulations with false positives to those with the correct number of rate shifts found. To remove the effect of lineage number, we compared groups of trees with the same number of diversification shifts. To account for a potential effect of tree imbalance, we compared the variance in lineage sizes inside trees, with or without false positives. There is no effect in most cases, except in the simulations with 4 or 5 rate shifts (p-values: $0.01$ and $3.6 \cdot 10^{-3}$, respectively, Mann-Whitney test). Thus, simulations with lineages of similar size are more likely to yield false positives only when they include more than 4 rate shifts. We also compared the variance in lineage sizes between simulations for which we recovered the correct number of events against those for which we recovered too few events. For every possible number of lineages, we find significantly lower variance for simulations that were correctly assessed. Thus, we only see a slight effect of the lineage size on the occurrence of false positives, whereas high variance in lineage size significantly increases false negatives. This indicates on the one hand, a tendency to overestimate the number of shifts when lineages are comparable in size, and on the other hand, problems with Medusa for identifying diversification shifts specific to a low number of species, as showed in the first part.\\

\subsubsection{Impact of events violating the model}
\begin{figure}[h!]
\begin{center}
\includegraphics{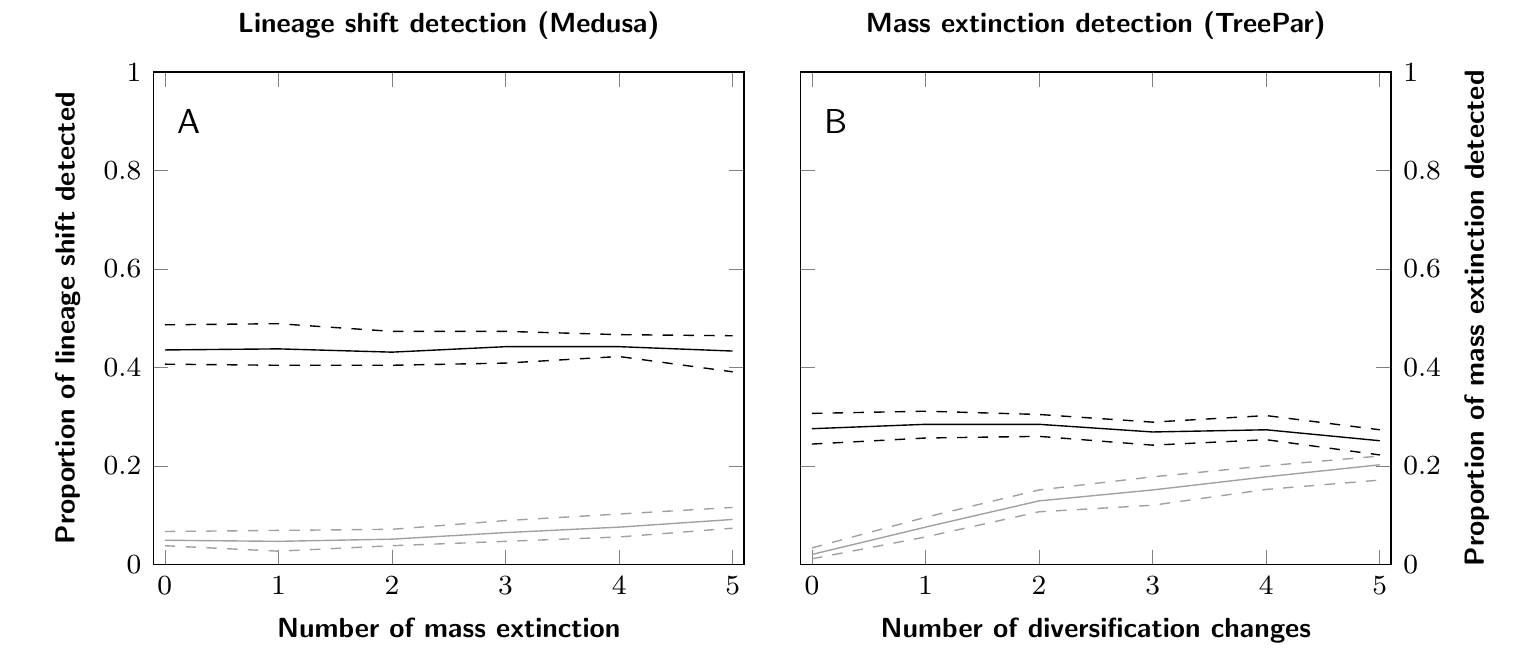}
\caption{Influence of increasing model violations on the tests. A: Lineage shift detection against an increasing number of mass extinctions; B: Mass extinction event detection against an increasing number of lineage shifts. Dark lines: simulations where the correct number of events was found, light lines: simulations where too many events was favoured.}
\label{events_opposed}
\end{center}
\end{figure}

\begin{figure}[h!]
\begin{center}
\includegraphics{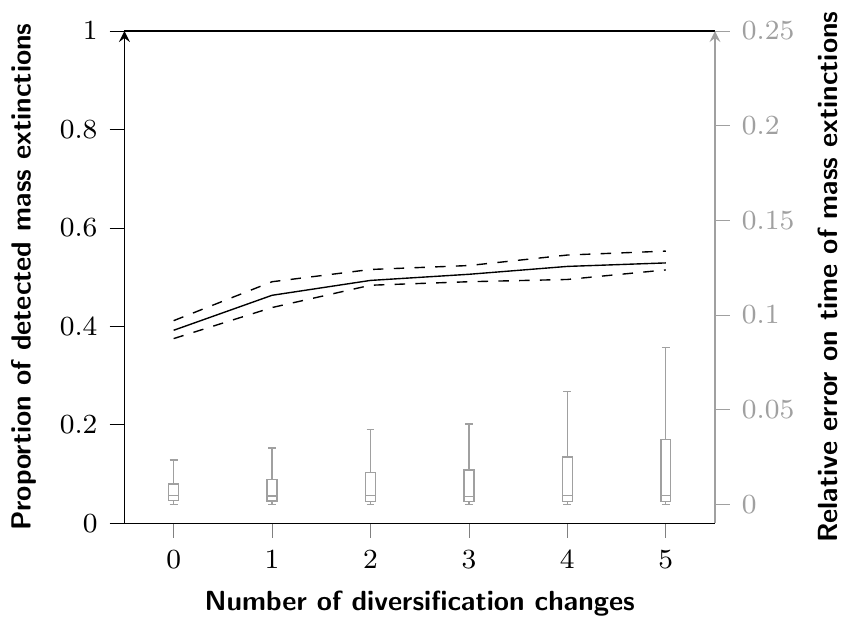}
  \end{center}
  \caption{Influence of the number of lineage shifts in a simulation upon the detection of mass extinctions. Line: proportion of detected mass extinctions; boxplots: distribution of the errors on their timing relative to the time of the first speciation event of the tree.}
  \label{events_opposed_distance}
\end{figure}

We tested the robustness of the methods by studying the behavior of (1) Medusa to detect rate shifts with an increasing number of mass extinctions, and (2) TreePar to detect mass extinction events with an increasing number of lineages shifts. The results of Medusa are unaffected by the number of mass extinctions in the simulations (Fig. \ref{events_opposed}). In contrast, an increase in the number of lineage shifts results in an increase of the proportion of false positives for TreePar ($2\%$ with no lineage shift vs. $20\%$ with five; Fig. \ref{events_opposed}). However, the accuracy of the estimate of the timing of the event is not affected (Fig. \ref{events_opposed_distance}). The number of lineage shifts has almost no impact on the probability of detecting a true mass extinction event, $i.e.$ on false negatives.\\

We note that false positive rates remain very low throughout all cases for Medusa, less than $10\%$ overall and around $5\%$ when dealing with simulations without mass extinctions (Fig. \ref{events_opposed}A). Recently, May $et$ $al.$ \cite{May2014} have also studied the performances of Medusa but with a different focus. Medusa also enables the characterization of diversification changes on incomplete phylogenies by letting the user assign species diversity at each tips of the tree. Two different equations are then used to calculate the likelihood function. One of them incorporates the likelihood of getting a specific number of species given a pair of $\lambda$ and $\mu$ after a certain amount of time, and is now used to account for the terminally input species numbers. May $et$ $al.$ simulated complete phylogenies before introducing uncertainties by sequentially collapsing some of the tips, and tested the different flavors of the three different Medusa algorithms ever made available. They found high Type I errors in every algorithm and biased parameter estimates. We note that in our study, we did not consider the estimation of the macro evolutionary parameters, and did not use unresolved trees, that can be used in Medusa to account uncertainties in the phylogeny. Interestingly, May $et$ $al.$ also tested the algorithm that we used in this study ($turboMedusa$, defined as $tMEDUSA$ in their study) on completely resolved trees, and found about the same rate of Type I errors as we did in the comparable trees (Figure S.20 of their study). Thus even though the focus of the two studies differs, they are in agreement in the few common analysis.\\

\subsubsection*{Impact of patterns of extinction} 
\begin{figure}[h!]
  \begin{center}
\makebox[\textwidth]{\includegraphics{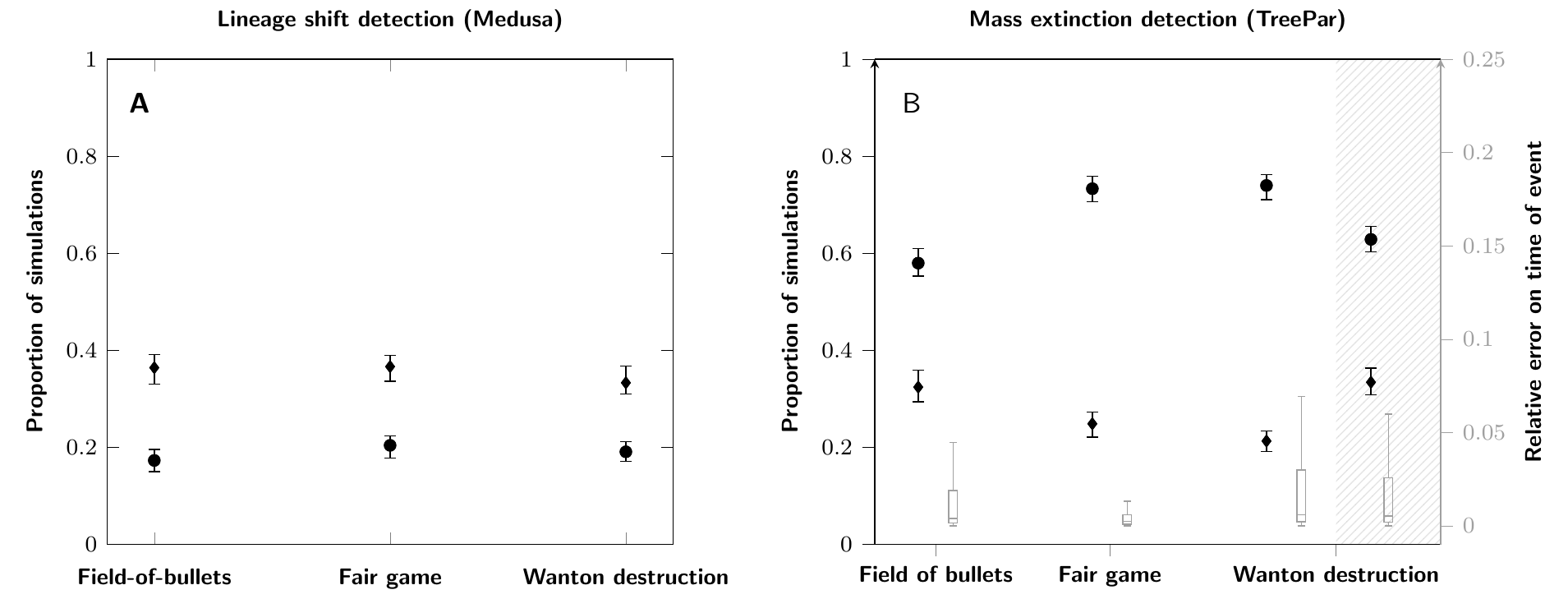}}
  \end{center}
\caption{Influence of distinct extinction scenarios on Medusa and TreePar predictions. A: Medusa outcome; diamonds: proportion of simulations where the model with the correct number of events is chosen; circles: proportion of simulations where a model with too many events is chosen; there are less correctly assessed simulations for Medusa because of the high number of lineage shifts in these simulations ($5$). B: TreePar outcome and error on the timing of events: boxplots: error on the timing of the estimated extinction relative to the first speciation event; blank background: detection of mass extinctions; dashed background: detection of global rate shifts; other symbols as in A.}
\label{mee_type}
\end{figure}

The effect of different scenarios of mass extinction on the results of Medusa and TreePar are presented in Figure \ref{mee_type}. First, as expected, no effect of the extinction scenarios is observed on the detection of lineage rate shifts detected by Medusa (Fig. \ref{mee_type}A). In contrast, the fair game and wanton destruction scenarios impact the estimation made by TreePar. They produce, for comparable levels of detection, more false positives than the field of bullets which was used in the previous simulations ($73\%$ and $74\%$ for fair and wanton against $58\%$ for field of bullets, Fig. \ref{mee_type}B). Irrespective of the type of mass extinction simulated, there are very few false negatives, $i.e.$ at least one extinction was detected in almost every tree. The error on the timing of this event was kept under $5\%$ of the root age. We also performed a search for global rate shifts in the case of wanton destruction (Fig. \ref{mee_type}B, dashed background). Regarding this scenario, we also compared simulations where all lineages undergo an increase of diversification after the mass extinction event against those who undergo a decrease and observe no difference between the outcomes of the two frameworks. Even though the shifts are different between lineages ($i.e.$, increase of diversification in some lineages, decrease in others), TreePar detects the period of this shift with more power than for the detection of the associated mass extinction ($34\%$ and $21\%$ correctly assessed simulations, respectively). Overall, these results show that departure from the simplest model of mass extinction should not affect our ability to detect these events in phylogenetic trees ($i.e.$ no increase in false negatives rate). But it should lead to an increase of false positive detections.\\

\section{Conclusion}

Previous studies involving mass extinctions and changes in macro-evolutionary rates have only focused on their effect on lineage through time plots \citep{Crisp2009}. This lead to the identification of a possible mass extinction event in some plants lineages around 32 Mya, which was further suggested to be linked with changes in climate. Recently, Hohna \cite{Hohna2013} developed a new algorithm to perform simulations with varying macro-evolutionary rates, allowing for mass extinction events. Other ongoing work aims at studying and simulating increasingly complex scenarios of diversification \citep{Hartmann2010, Morlon2010}, but we would like to emphasize that no method allows the simultaneous discovery of both time-specific or lineage-specific rate changes and mass extinction events.\\

The study of diversification rates has become a standard part of the analysis of large phylogenetic trees \citep{Meredith2011, Jetz2012, Near2013}, and recent efforts have also assessed the methods used when their assumptions are violated \citep{Rabosky2014a}. We have shown that departure from the assumption of consistency in rates across lineages causes a large increase in false positives when looking for mass extinction events. This can be problematic as we know that rate consistency rarely holds \citep{Rabosky2007a, Jetz2012, Barker2013}, and casts doubts on our ability to reliably find such events using only phylogenetic trees. Nevertheless, an increasing number of disparities between lineages caused neither a decrease in the probability of detecting an event nor an increase in the error on its timing. As we observed the same pattern under more complex scenarios of extinction, the difficulty in detecting the K-Pg event in mammals is therefore probably not due to biases in the methods used. We might be limited by the power of TreePar to detect mass extinction events, although in simulations we reach 60\% of true events detected for a tree size similar to that of mammals.\\

Recent efforts aim to reach a better agreement between paleontological and molecular data \citep{Morlon2011}, including looking for mass extinctions in molecular phylogenies. For instance, there is much debate on whether the K-Pg extinction event triggered the mammalian diversification \citep{Bininda-Emonds2007, Meredith2011, Stadler2011, DosReis2012, OLeary2013}. The fossil record also indicates higher extinction rates of mammalians species around $65$ Mya \citep{Wilson2005}. In this work, we have shown that for phylogenetic trees similar in size to that of mammals (i.e. $ca.$ $5000$ species), the signal for mass extinctions was usually recovered in the tree, even though lineage discrepancies in macro-evolutionary rates had a tendency to yield more false positives. Thus, if the ancestor lineages of the extant mammal families did experience a mass extinction at the K-Pg boundary, we should theoretically be able to identify it using phylogenetic trees. The underlying assumption about the mass extinction made when using TreePar is that lineages are terminated randomly with a fixed $\rho$ value everywhere in the tree, $i.e.$ a field of bullets type of mass extinction. But other models of extinction seem to increase false positives but not false negatives, not explaining difficulties in finding a K-Pg signal in real phylogenetic trees.\\

Recent studies have used Markov processes to account for the effect of specific traits upon the probability of extinction of a species, thus extending models of mass extinction beyond the field of bullets \citep{Faller2012}. Such models can be used for instance to estimate the loss of phylogenetic diversity after a mass extinction event \citep{Lambert2013}. Our simulations can be seen as a special case of such models, where the trait influencing survival probabilities is the diversification value of the species. We have shown that more complex models of mass extinction cause more false positive detection than the simple field of bullets, as well as a decrease in the error for the fair game scenario. Choosing a specific model of extinction (field of bullets, wanton destruction, fair game) might require the incorporation of fossil information into the phylogenetic tree, and thus the further development of methods capable of dealing with both molecular and fossil data. \\


\section{Competing interests}
The authors declare that they have no competing interests.

\section{Authors contributions}
SL, MRR and NS designed the study, SL performed the simulations, SL, MRR and NS analyzed the results and wrote the manuscript.

\section{Acknowledgements}
This work was supported by the ProDoc grant number 134931 of the Swiss National Science Fondation; and État de Vaud. The computations were performed at the Vital-IT (http://www.vital-it.ch) Center for high-performance computing of the SIB Swiss Institute of Bioinformatics. We thank Tanja Stadler, Daniele Silvestro and four anonymous reviewers for helpful discussions.
\bibliographystyle{humannat} 
\renewcommand\bibname{Reference}

\bibliography{/home/sacha/Documents/svn/library}      

\begin{thebibliography}{}

\bibitem[\protect\astroncite{Alfaro et~al.}{2009}]{Alfaro2009}
Alfaro, M.~E., F.~Santini, C.~Brock, H.~Alamillo, A.~Dornburg, D.~L. Rabosky,
  G.~Carnevale, and L.~J. Harmon\leavevmode\nopagebreak\newline 2009.
\newblock {Nine exceptional radiations plus high turnover explain species
  diversity in jawed vertebrates.}
\newblock {\em Proceedings of the National Academy of Sciences of the United
  States of America}, 106(32):13410--13414.

\bibitem[\protect\astroncite{Barker et~al.}{2013}]{Barker2013}
Barker, F.~K., K.~J. Burns, J.~Klicka, S.~M. Lanyon, and I.~J.
  Lovette\leavevmode\nopagebreak\newline 2013.
\newblock {Going to extremes: contrasting rates of diversification in a recent
  radiation of new world passerine birds.}
\newblock {\em Systematic biology}, 62(2):298--320.

\bibitem[\protect\astroncite{Benton}{1995}]{Benton1995}
Benton, M.~J.\leavevmode\nopagebreak\newline 1995.
\newblock {Diversification and Extinction in the History of Life}.
\newblock {\em Science}, 268(5207):52--58.

\bibitem[\protect\astroncite{Bininda-Emonds et~al.}{2007}]{Bininda-Emonds2007}
Bininda-Emonds, O. R.~P., M.~Cardillo, K.~E. Jones, R.~D.~E. MacPhee, R.~M.~D.
  Beck, R.~Grenyer, S.~a. Price, R.~a. Vos, J.~L. Gittleman, and
  A.~Purvis\leavevmode\nopagebreak\newline 2007.
\newblock {The delayed rise of present-day mammals.}
\newblock {\em Nature}, 446(7135):507--512.

\bibitem[\protect\astroncite{Condamine et~al.}{2013}]{Condamine2013}
Condamine, F.~L., J.~Rolland, and H.~Morlon\leavevmode\nopagebreak\newline
  2013.
\newblock {Macroevolutionary perspectives to environmental change.}
\newblock {\em Ecology letters}, 16 Suppl 1:72--85.

\bibitem[\protect\astroncite{Crisp and Cook}{2009}]{Crisp2009}
Crisp, M.~D. and L.~G. Cook\leavevmode\nopagebreak\newline 2009.
\newblock {Explosive radiation or cryptic mass extinction? Interpreting
  signatures in molecular phylogenies.}
\newblock {\em Evolution}, 63(9):2257--65.

\bibitem[\protect\astroncite{{Dos Reis} et~al.}{2012}]{DosReis2012}
{Dos Reis}, M., J.~Inoue, M.~Hasegawa, R.~J. Asher, P.~C.~J. Donoghue, and
  Z.~Yang\leavevmode\nopagebreak\newline 2012.
\newblock {Phylogenomic datasets provide both precision and accuracy in
  estimating the timescale of placental mammal phylogeny.}
\newblock {\em Proceedings. Biological sciences / The Royal Society},
  279(1742):3491--500.

\bibitem[\protect\astroncite{Eble}{1999}]{Eble1999}
Eble, G.~J.\leavevmode\nopagebreak\newline 1999.
\newblock {Paleontological Society On the Dual Nature of Chance in Evolutionary
  Biology and Paleobiology On the dual nature of chance in evolutionary biology
  and paleobiology}.
\newblock {\em Paleobiology}, 25(1):75--87.

\bibitem[\protect\astroncite{Erwin}{2006}]{Erwin2006}
Erwin, D.~H.\leavevmode\nopagebreak\newline 2006.
\newblock {\em {Extinction: How Life on Earth Nearly Ended 250 Million Years
  Ago}}.
\newblock Princeton: Princeton University Press.

\bibitem[\protect\astroncite{Etienne et~al.}{2012}]{Etienne2012}
Etienne, R.~S., B.~Haegeman, T.~Stadler, T.~Aze, P.~N. Pearson, A.~Purvis, and
  A.~B. Phillimore\leavevmode\nopagebreak\newline 2012.
\newblock {Diversity-dependence brings molecular phylogenies closer to
  agreement with the fossil record.}
\newblock {\em Proceedings. Biological sciences / The Royal Society},
  279(1732):1300--9.

\bibitem[\protect\astroncite{Faller et~al.}{2008}]{Faller2008}
Faller, B., F.~Pardi, and M.~Steel\leavevmode\nopagebreak\newline 2008.
\newblock {Distribution of phylogenetic diversity under random extinction.}
\newblock {\em Journal of theoretical biology}, 251(2):286--96.

\bibitem[\protect\astroncite{Faller and Steel}{2012}]{Faller2012}
Faller, B. and M.~Steel\leavevmode\nopagebreak\newline 2012.
\newblock {Trait-Dependent Extinction Leads to Greater Expected Biodiversity
  Loss}.
\newblock {\em SIAM Journal on Discrete Mathematics}, 26(2):472--481.

\bibitem[\protect\astroncite{FitzJohn et~al.}{2009}]{FitzJohn2009}
FitzJohn, R.~G., W.~P. Maddison, and S.~P. Otto\leavevmode\nopagebreak\newline
  2009.
\newblock {Estimating trait-dependent speciation and extinction rates from
  incompletely resolved phylogenies.}
\newblock {\em Systematic biology}, 58(6):595--611.

\bibitem[\protect\astroncite{Hartmann et~al.}{2010}]{Hartmann2010}
Hartmann, K., D.~Wong, and T.~Stadler\leavevmode\nopagebreak\newline 2010.
\newblock {Sampling trees from evolutionary models.}
\newblock {\em Systematic biology}, 59(4):465--76.

\bibitem[\protect\astroncite{H\"{o}hna}{2013}]{Hohna2013}
H\"{o}hna, S.\leavevmode\nopagebreak\newline 2013.
\newblock {Fast simulation of reconstructed phylogenies under global
  time-dependent birth-death processes.}
\newblock {\em Bioinformatics}, 29(11):1367--74.

\bibitem[\protect\astroncite{Jetz et~al.}{2012}]{Jetz2012}
Jetz, W., G.~H. Thomas, J.~B. Joy, K.~Hartmann, and a.~O.
  Mooers\leavevmode\nopagebreak\newline 2012.
\newblock {The global diversity of birds in space and time.}
\newblock {\em Nature}, 491(7424):444--8.

\bibitem[\protect\astroncite{Lambert and Steel}{2013}]{Lambert2013}
Lambert, A. and M.~Steel\leavevmode\nopagebreak\newline 2013.
\newblock {Predicting the loss of phylogenetic diversity under non-stationary
  diversification models}.
\newblock {\em Journal of Theoretical Biology}, 337:111--24.

\bibitem[\protect\astroncite{Maddison et~al.}{2007}]{Maddison2007}
Maddison, W.~P., P.~E. Midford, and S.~P. Otto\leavevmode\nopagebreak\newline
  2007.
\newblock {Estimating a binary character's effect on speciation and
  extinction.}
\newblock {\em Systematic biology}, 56(5):701--10.

\bibitem[\protect\astroncite{May and Moore}{2014}]{May2014}
May, M.~R. and B.~R. Moore\leavevmode\nopagebreak\newline 2014.
\newblock {How Well Can We Detect Shifts in Rates of Lineage Diversification ?
  A Simulation Study of Sequential AIC Methods}.
\newblock {\em bioRxiv}.

\bibitem[\protect\astroncite{Mayrose et~al.}{2011}]{Mayrose2011}
Mayrose, I., S.~H. Zhan, C.~J. Rothfels, K.~Magnuson-Ford, M.~S. Barker, L.~H.
  Rieseberg, and S.~P. Otto\leavevmode\nopagebreak\newline 2011.
\newblock {Recently formed polyploid plants diversify at lower rates.}
\newblock {\em Science}, 333(6047):1257.

\bibitem[\protect\astroncite{Meredith et~al.}{2011}]{Meredith2011}
Meredith, R.~W., J.~E. Jane\v{c}ka, J.~Gatesy, O.~a. Ryder, C.~a. Fisher, E.~C.
  Teeling, A.~Goodbla, E.~Eizirik, T.~L.~L. Sim\~{a}o, T.~Stadler, D.~L.
  Rabosky, R.~L. Honeycutt, J.~J. Flynn, C.~M. Ingram, C.~Steiner, T.~L.
  Williams, T.~J. Robinson, A.~Burk-Herrick, M.~Westerman, N.~a. Ayoub, M.~S.
  Springer, and W.~J. Murphy\leavevmode\nopagebreak\newline 2011.
\newblock {Impacts of the Cretaceous Terrestrial Revolution and KPg extinction
  on mammal diversification.}
\newblock {\em Science}, 334(6055):521--4.

\bibitem[\protect\astroncite{Morlon et~al.}{2011}]{Morlon2011}
Morlon, H., T.~L. Parsons, and J.~B. Plotkin\leavevmode\nopagebreak\newline
  2011.
\newblock {Reconciling molecular phylogenies with the fossil record}.
\newblock {\em Proceedings of the National Academy of Sciences},
  108(39):16327--32.

\bibitem[\protect\astroncite{Morlon et~al.}{2010}]{Morlon2010}
Morlon, H., M.~D. Potts, and J.~B. Plotkin\leavevmode\nopagebreak\newline 2010.
\newblock {Inferring the dynamics of diversification: a coalescent approach.}
\newblock {\em PLoS biology}, 8(9).

\bibitem[\protect\astroncite{Near et~al.}{2013}]{Near2013}
Near, T.~J., A.~Dornburg, R.~I. Eytan, B.~P. Keck, W.~L. Smith, K.~L. Kuhn,
  J.~A. Moore, S.~A. Price, F.~T. Burbrink, M.~Friedman, and P.~C.
  Wainwright\leavevmode\nopagebreak\newline 2013.
\newblock {Phylogeny and tempo of diversification in the superradiation of
  spiny-rayed fishes}.
\newblock {\em Proceedings of the National Academy of Sciences},
  110(31):12738--43.

\bibitem[\protect\astroncite{Nee}{1997}]{Nee1997}
Nee, S.\leavevmode\nopagebreak\newline 1997.
\newblock {Extinction and the Loss of Evolutionary History}.
\newblock {\em Science}, 278(5338):692--694.

\bibitem[\protect\astroncite{Nee et~al.}{1994}]{Nee1994}
Nee, S., E.~C. Holmes, R.~May, and P.~Harvey\leavevmode\nopagebreak\newline
  1994.
\newblock {Extinction rates can be estimated from molecular phylogenies.}
\newblock {\em Philosophical transactions of the Royal Society of London.
  Series B, Biological sciences}, 344(1307):77--82.

\bibitem[\protect\astroncite{O'Leary et~al.}{2013}]{OLeary2013}
O'Leary, M.~a., J.~I. Bloch, J.~J. Flynn, T.~J. Gaudin, A.~Giallombardo, N.~P.
  Giannini, S.~L. Goldberg, B.~P. Kraatz, Z.-X. Luo, J.~Meng, X.~Ni, M.~J.
  Novacek, F.~a. Perini, Z.~S. Randall, G.~W. Rougier, E.~J. Sargis, M.~T.
  Silcox, N.~B. Simmons, M.~Spaulding, P.~M. Velazco, M.~Weksler, J.~R. Wible,
  and A.~L. Cirranello\leavevmode\nopagebreak\newline 2013.
\newblock {The placental mammal ancestor and the post-K-Pg radiation of
  placentals.}
\newblock {\em Science}, 339(6120):662--7.

\bibitem[\protect\astroncite{{R Core Team}}{2013}]{R}
{R Core Team}\leavevmode\nopagebreak\newline 2013.
\newblock {\em {R: A Language and Environment for Statistical Computing}}.
\newblock R Foundation for Statistical Computing, Vienna, Austria.

\bibitem[\protect\astroncite{Rabosky}{2014}]{Rabosky2014a}
Rabosky, D.~L.\leavevmode\nopagebreak\newline 2014.
\newblock {Automatic Detection of Key Innovations, Rate Shifts, and
  Diversity-Dependence on Phylogenetic Trees}.
\newblock {\em PLoS ONE}, 9(2):e89543.

\bibitem[\protect\astroncite{Rabosky et~al.}{2007}]{Rabosky2007a}
Rabosky, D.~L., S.~C. Donnellan, A.~L. Talaba, and I.~J.
  Lovette\leavevmode\nopagebreak\newline 2007.
\newblock {Exceptional among-lineage variation in diversification rates during
  the radiation of Australia's most diverse vertebrate clade.}
\newblock {\em Proceedings. Biological sciences / The Royal Society},
  274(1628):2915--23.

\bibitem[\protect\astroncite{Raup}{1992}]{Raup1992}
Raup, D.~M.\leavevmode\nopagebreak\newline 1992.
\newblock {\em {Extinction: Bad Genes or Bad Luck ?}}
\newblock New York: W. W. Norton \& Company.

\bibitem[\protect\astroncite{Raup and Sepkoski}{1982}]{Raup1982}
Raup, D.~M. and J.~J. Sepkoski\leavevmode\nopagebreak\newline 1982.
\newblock {Mass Extinction in the Marine Fossil Record}.
\newblock {\em Science}, 215(4539):1501--1503.

\bibitem[\protect\astroncite{Ricklefs}{2007}]{Ricklefs2007}
Ricklefs, R.~E.\leavevmode\nopagebreak\newline 2007.
\newblock {Estimating diversification rates from phylogenetic information.}
\newblock {\em Trends in ecology \& evolution}, 22(11):601--10.

\bibitem[\protect\astroncite{Silvestro et~al.}{2011}]{Silvestro2011}
Silvestro, D., J.~Schnitzler, and G.~Zizka\leavevmode\nopagebreak\newline 2011.
\newblock {A Bayesian framework to estimate diversification rates and their
  variation through time and space.}
\newblock {\em BMC evolutionary biology}, 11(1):311.

\bibitem[\protect\astroncite{Stadler}{2011a}]{Stadler2011}
Stadler, T.\leavevmode\nopagebreak\newline 2011a.
\newblock {Mammalian phylogeny reveals recent diversification rate shifts.}
\newblock {\em Proceedings of the National Academy of Sciences of the United
  States of America}, 108(15):6187--92.

\bibitem[\protect\astroncite{Stadler}{2011b}]{Stadler2011d}
Stadler, T.\leavevmode\nopagebreak\newline 2011b.
\newblock {Simulating trees with a fixed number of extant species.}
\newblock {\em Systematic biology}, 60(5):676--84.

\bibitem[\protect\astroncite{Stadler and Bokma}{2013}]{Stadler2013}
Stadler, T. and F.~Bokma\leavevmode\nopagebreak\newline 2013.
\newblock {Estimating speciation and extinction rates for phylogenies of higher
  taxa.}
\newblock {\em Systematic biology}, 62(2):220--30.

\bibitem[\protect\astroncite{Wilson}{2005}]{Wilson2005}
Wilson, G.~P.\leavevmode\nopagebreak\newline 2005.
\newblock {Mammalian Faunal Dynamics During the Last 1.8 Million Years of the
  Cretaceous in Garfield County, Montana}.
\newblock {\em Journal of Mammalian Evolution}, 12(1-2):53--76.

\end{thebibliography}






\section*{Supplementary material}

# Detecting patterns of species diversification in the presence of both rate shifts and mass extinctions

Sacha Laurent, Marc Robinson-Rechavi and Nicolas Salamin

Department of Ecology and Evolution, University of Lausanne, Biophore, 1015, Lausanne, Switzerland
Swiss Institute of Bioinformatics, Quartier Sorge, 1015 Lausanne, Switzerland

We compared our backward and forward-like algorithms to assess if the backward approach could introduce a potential bias in the branching times distribution. This could then have an effect on the rate shifts that are inferred by Medusa. We therefore did a series of simulations to test the differences between the two algorithms.

The Supplemental Figure 1 show that for every species-numbered tree, forward and backward algorithm yields comparable branching times densities. As both methods tested in our paper take branching times as only input, we conclude that the way we simulated our trees in our paper as little chance to bias the results in anyway.

The Supplemental Figure 2 shows the results for 50 trees of 500 and 200 species. We observe that in both cases, similar numbers of diversification shifts were found by Medusa. Regarding distances between simulated and identified shifts by Medusa, most of the events found were situated at a close distance of the real event, even though we observe a slight tendency for a lower precision in the forward-like algorithm. This could be an effect of the overall size of the tree (as the forward-like would always be smaller than the 200 or 500 species sharp backward tree). Indeed, as we see in the Fig 3 of the manuscript, the effect of the overall tree size is low in general, but there is a slight difference for small clades and 200/500 species trees, that could explain the discrepancy we are seeing here. In general, we feel that these new simulations prove the small effect that our simulation schemes has in distorting the main results of our paper.

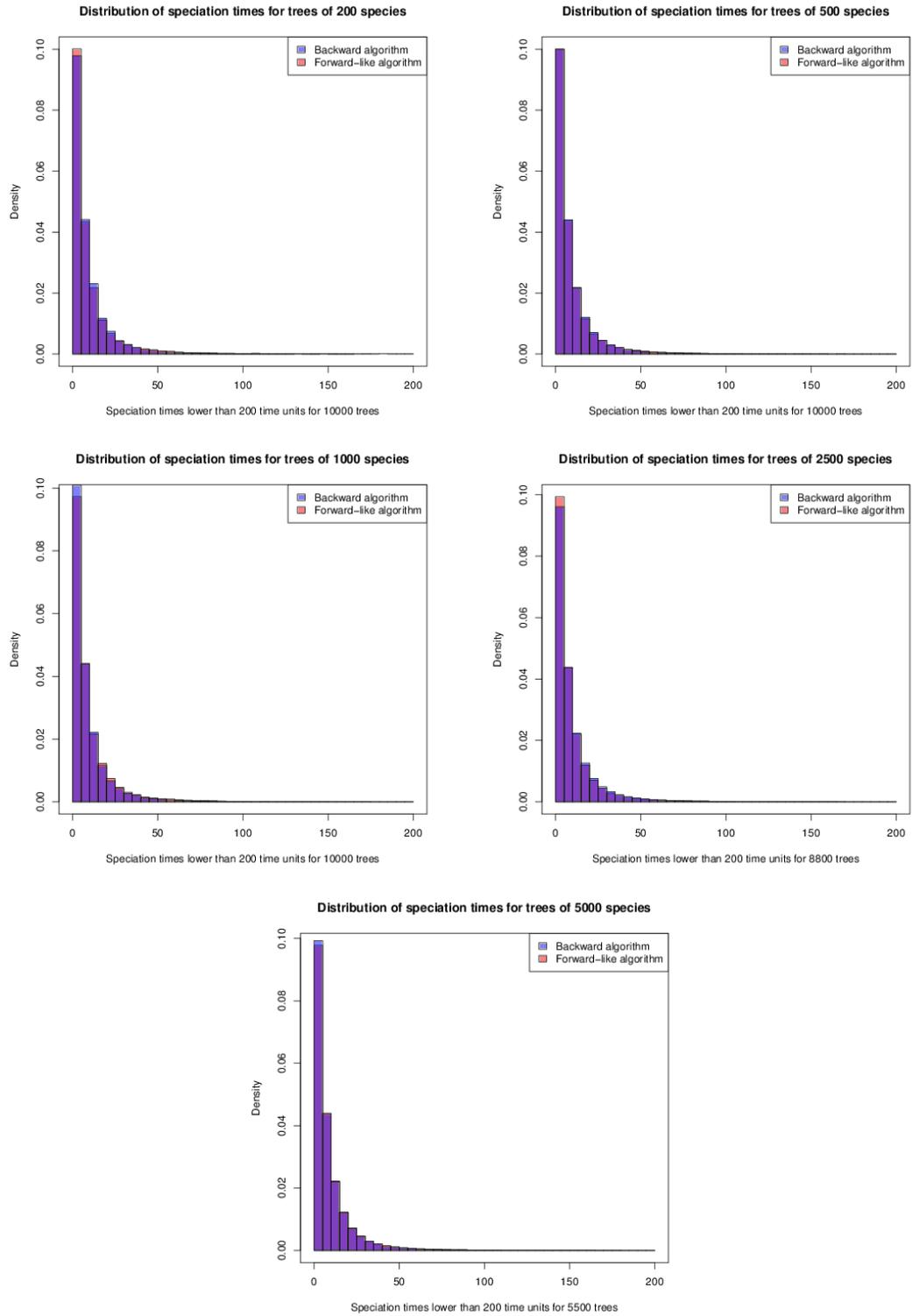

Suppl. Figure 1: Comparison of our backward and forward-like algorithms with trees of different sizes when analysed with Medusa. We simulated 10000 trees of size 200, 500, 1000, 2500 and 5000 tips.

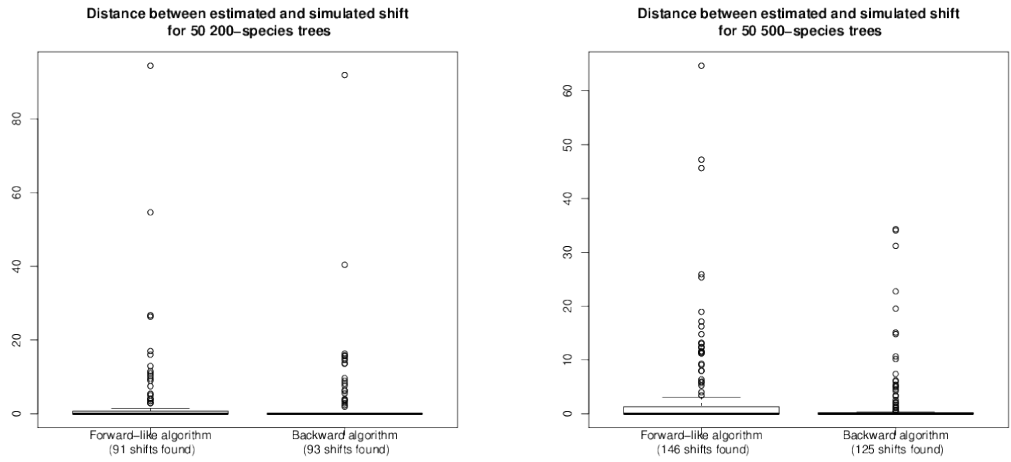

Suppl. Figure 2: Differences in branching times between trees obtained with our backward or the forward-like algorithms. We simulated 50 trees with either 200 (left panel) or 500 (right panel) tips.


\end{document}